\documentclass[aip,reprint,rsi]{revtex4-1}

\usepackage{graphicx}
\usepackage{dcolumn}
\usepackage{bm}
\usepackage{color}

\begin{document}

\preprint{AIP/123-QED}

\title[Zeeman slower made simple]{Zeeman slowers made simple with permanent magnets in a Halbach configuration}

\author{P.~Cheiney}
\author{O.~Carraz}
\affiliation{Universit\'e de Toulouse, UPS - 118 route de Narbonne, F-31062 Toulouse, France.}
\affiliation{CNRS, UMR 5589, Laboratoire Collisions Agr\'egats R\'eactivit\'e, IRSAMC - F-31062 Toulouse, France.}
\author{D.~Bartoszek-Bober}\affiliation{Marian Smoluchowski Institute of Physics, Jagiellonian University, Cracow, Poland.}
\author{S.~Faure}
\author{F.~Vermersch}
\author{C.~M.~Fabre}
\author{G.~L.~Gattobigio}
\author{T.~Lahaye}
\author{D.~Gu\'ery-Odelin}
\author{R.~Mathevet}
\email{renaud.mathevet@irsamc.ups-tlse.fr}
\affiliation{Universit\'e de Toulouse, UPS - 118 route de Narbonne, F-31062 Toulouse, France.}
\affiliation{CNRS, UMR 5589, Laboratoire Collisions Agr\'egats R\'eactivit\'e, IRSAMC - F-31062 Toulouse, France.}

\date{\today}

\begin{abstract}
We describe a simple Zeeman slower design using permanent magnets. Contrary to common wire-wound setups no electric power and water cooling are required. In addition, the whole system can be assembled and disassembled at will. The magnetic field is however transverse to the atomic motion and an extra repumper laser is necessary. A Halbach configuration of the magnets produces a high quality magnetic field and no further adjustment is needed. After optimization of the laser parameters, the apparatus produces an intense beam of slow and cold $^{87}$Rb atoms. With typical fluxes of $1$ to $5\times10^{10}~\mathrm{atoms/s}$ at $30~\mathrm{m\cdot s^{-1}}$, our apparatus efficiently loads a large magneto-optical trap with more than $10^{10}$ atoms in one second, which is an ideal starting point for degenerate quantum gases experiments.
\end{abstract}

\maketitle

\section{\label{sec:Introduction} Introduction}

Nowadays, many atomic physics experiments study or use quantum degenerate gases for which a large initial sample of cold atoms is required. A wide variety of experimental techniques has been developed for slowing and cooling atoms. Many of them rely on the radiation pressure from quasi resonant light. In particular, since their first realization,\cite{Phillips1982} Zeeman slowers have become very popular for loading magneto-optical traps (MOT). These cold atoms reservoirs are then an ideal starting point to implement other techniques for further cooling.

Recently, several Zeeman slowers using permanent magnets have been built\footnote{Y. Ovchinnikov, J. McCelland, D. Comparat, G. Reinaudi, private communications} following the proposal of Ref.~\onlinecite{Ovchinnikov2007} (see also Ref.~\onlinecite{Bagayev2001} for a somewhat different approach formerly used). Here, we present an alternative design based on a Halbach configuration\cite{Halbach1980} of the magnets and demonstrate fully satisfactory operation. Before going into details, let us emphasize some advantages of the setup:
 \begin{itemize}
   \item simple to implement, compact and light,\footnote{$1200~\mathrm{mm}$ long, $120\times130~\mathrm{mm^2}$ cross section; Dural holders, magnets and shield weights are respectively $5.0$, $2.5$ and $4.5~\mathrm{kg}$. It takes half a day of work to put parts together once machining is done.}
   \item no electric power consumption nor water cooling,
   \item high fields with excellent transverse homogeneity,
   \item very smooth longitudinal profile and low stray magnetic fields,
   \item easy to assemble and disassemble without vacuum breaking e.g. for high-temperature baking out.
 \end{itemize}

This paper is organized as follows. In the next section, we first give the basics of the theoretical framework and then compare our permanent magnets approach with the usual wire-wound technique. Then we collect in Sec.~\ref{sec:Field:calculations} some information on magnets, shields, field calculations and measurements useful to characterize our setup described in Sec.~\ref{sec:Mechanics:and:measurements}. We subsequently detail in Sec.~\ref{sec:Experimental:apparatus} the whole experimental apparatus before we finally present the Zeeman slower performances in Sec.~\ref{sec:Zeeman:Slower:performances}.

\section{\label{sec:Zeeman:slowers:designs}Zeeman slowers designs}
    \subsection{\label{subsec:Zeeman:Theory:specifications}Notations and field specifications}
In a Zeeman slower, atoms are decelerated by scattering photons from a near resonant counter propagating laser. Let $Oz$ denote the mean atom and light propagation axis, $\Gamma$ and $\mu$ the linewidth and magnetic moment of the atomic transition, $\bm{k}$ the light wave vector, $m$ the atomic mass and $v(z)$ the velocity at $z$ of an atom entering the field at $z=0$. To keep atoms on resonance, changes in the Doppler shift $kv(z)$ are compensated for by opposite changes of the Zeeman effect $\mu B(z)$ in an inhomogeneous magnetic field $B(z)$.\cite{Phillips1998} We use an {\em increasing field} configuration\cite{barrett1991} for better performance with $^{87}$Rb.

As the scattering rate cannot exceed $\Gamma/2$, the maximum achievable acceleration is:
\begin{equation}
\label{Eqn:amax}
a_{\mathrm{max}}=\frac{\Gamma}{2}\frac{\hbar k}{m}.\nonumber
\end{equation}
To keep a safety margin, the ideal magnetic field profile $B_{\mathrm{id}}(z)$ is calculated for only a fraction $\eta=0.75$ of $a_{\mathrm{max}}$. Energy conservation reads $v(z)^2=v(0)^2-2\eta a_{\mathrm{max}}z$ so that:
\begin{equation}
\label{Eqn:bid}
B_{\mathrm{id}}(z)=B_{\mathrm{bias}}+\Delta B\left(1-\sqrt{1-z/L}\right),
\end{equation}
 where the length of the apparatus is $L=v(0)^2/2\eta a_{\mathrm{max}}$ and $\mu \Delta B/\hbar=k v(0)$ assuming $v(L)\ll v(0)$. $v(0)$ defines the capture velocity as, in principle, all velocity classes below $v(0)$ are slowed down to $v(L)$. A bias field $B_{\mathrm{bias}}$ is added for technical reasons discussed later on (Sec.~\ref{subsubsec:Mechanics:and:field:characterization:field:parameters}). To match the resonance condition, lasers must be detuned from the atomic transition by:
\begin{equation}
\label{Eqn:delta0}
\delta_0\approx\mu(B_{\mathrm{bias}}+\Delta B)/h.
\end{equation}

Finally, slowing must be efficient over the whole atomic beam diameter. A conservative estimate of the acceptable field variations in a cross section is $\delta B=h\Gamma/|\mu|$ which amounts to $\sim4~\mathrm{G}$ given the rubidium linewidth $\Gamma\approx2\pi\times6~\mathrm{MHz}$. Here, such high transverse homogeneity, intrinsic to solenoids, is achieved using permanent magnets in a particular geometry inspired by Halbach cylinders. This represents a major improvement with respect to the original proposal\cite{Ovchinnikov2007} which the reader is also referred to for a more detailed theoretical analysis.

    \subsection{\label{subsec:Zeeman:Implementations}Different implementations}
        \subsubsection{\label{subsubsec:Zeeman:wire:wound:permanent:magnets} Wire-wound vs permanent magnets slowers}

\begin{figure}
\includegraphics[width=\columnwidth]{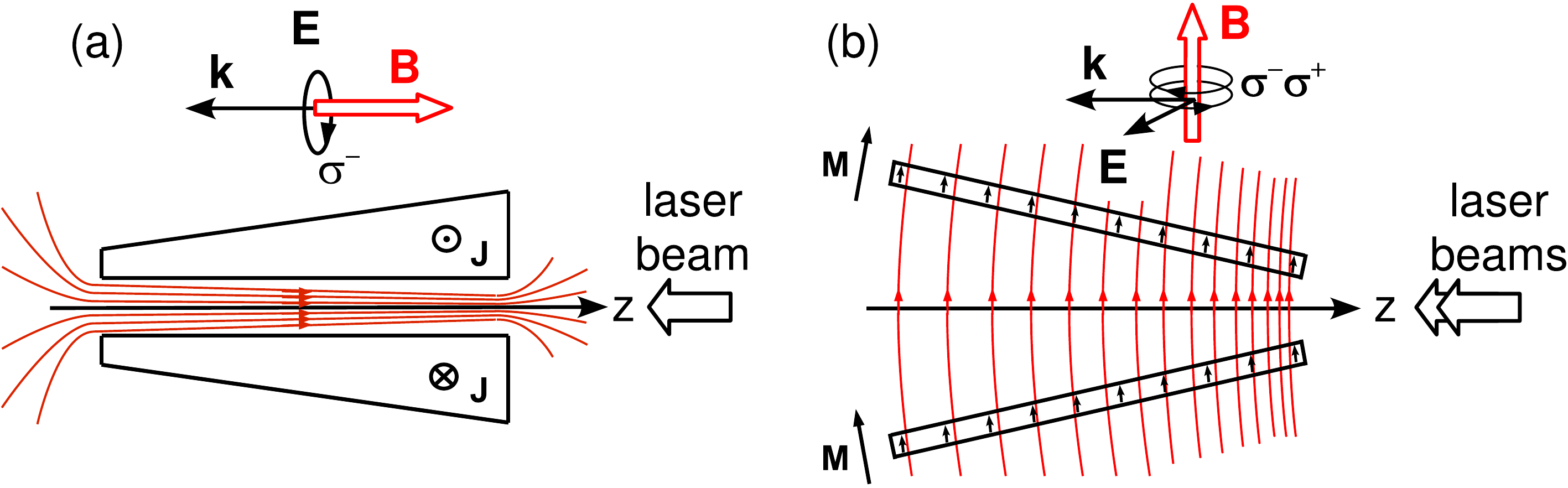}
\caption{Color online. Zeeman slower configurations. (a) Conventional wire-wound tapered solenoid; magnetic field is longitudinal, $\sigma^-$ light is used. $\bm{J}$ denotes the current density vector. (b) Use of long tilted permanent magnets; magnetic field is transverse. Light polarisation is decomposed in its $\sigma^\pm$ components and a repumper is needed. $\bm{M}$ denotes the magnetization of the material.\label{Fig:longitrans}}
\end{figure}

In most Zeeman slowers the magnetic field is generated with current flowing in wires wound around the atomic beam. The ideal profile of Eq.~(\ref{Eqn:bid}) is commonly obtained varying the number of layers (Fig.~\ref{Fig:longitrans}~(a)) or more recently the winding pitch.\cite{bell2010} The field is then essentially that of a solenoid: longitudinal and very homogeneous in a transverse plane. There are usually some drawbacks to this technique. Winding of up to several tens of layers has to be done with care to get a smooth longitudinal profile. It represents hundreds of meters and typically ten kilograms of copper wire so the construction can be somewhat tedious. It is moreover done once for all and cannot be removed later on. As a result, only moderate baking out is possible which may limit vacuum quality. Finally, electric power consumption commonly amounts to hundreds of watts so water cooling can be necessary.

Of course, the use of permanent magnets circumvents these weak points. In the original proposal,\cite{Ovchinnikov2007} two rows of centimeter-sized magnets are positioned on both sides of the atomic beam. Contrary to wire-wound systems the field is thus transverse.\footnote{we could not find a configuration that produces a longitudinal magnetic field of several hundreds of Gauss with a reasonable amount of magnetic material.} Fortunately, slowing in such a configuration is also possible.\cite{melentiev2004} Although this initial design is very simple, the field varies quickly off axis, typically several tens of Gauss over the beam diameter, which may reduce the slower efficiency.

        \subsubsection{\label{subsubsec:Zeeman:Halbach:configuration} Halbach configuration}
\begin{figure}
\includegraphics[width=\columnwidth]{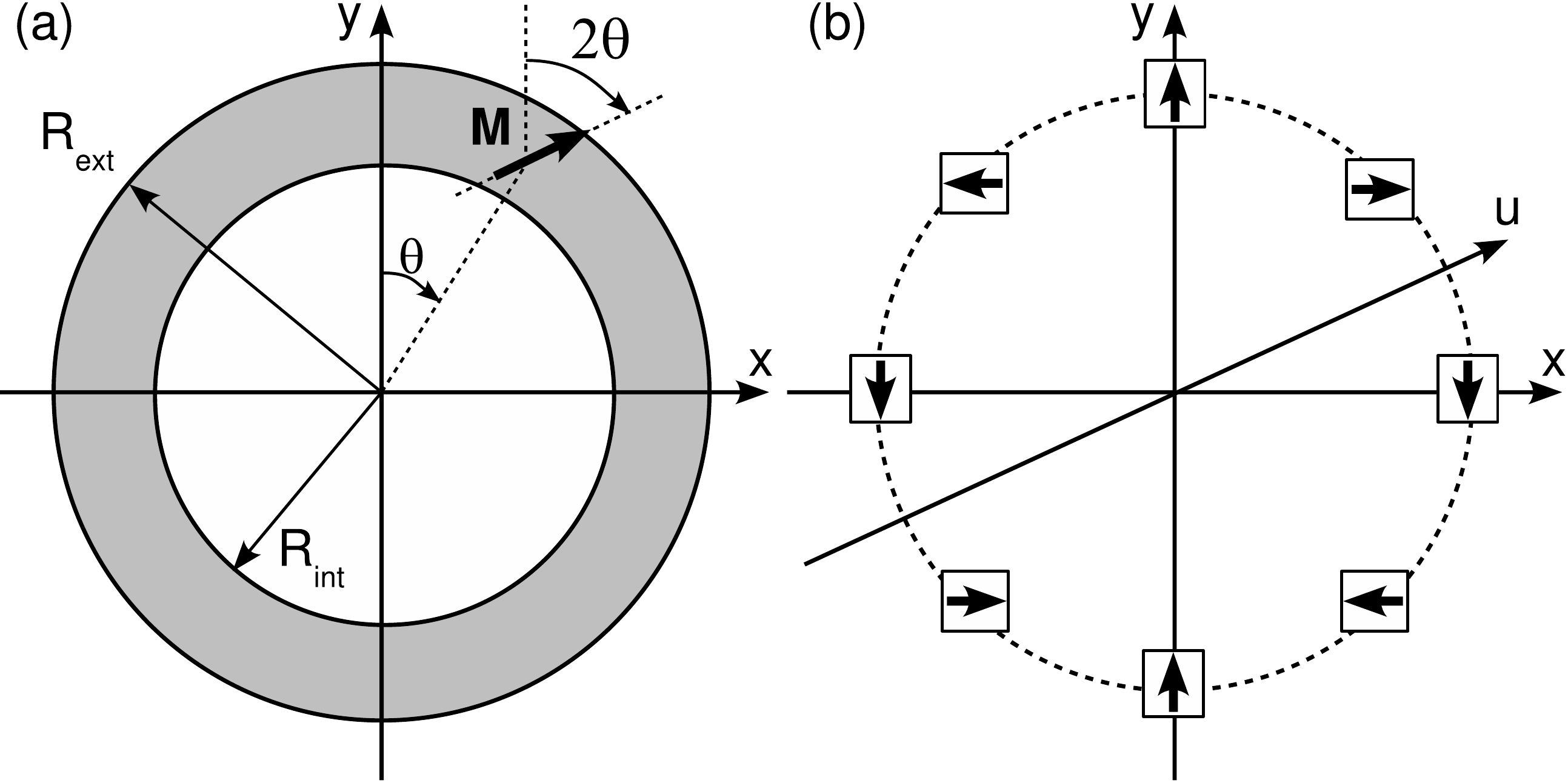}
\caption{(a) Notations for Halbach cylinder. (b) Transverse cross section showing a 8-pole Halbach configuration.\label{Fig:Halbach}}
\end{figure}
A way to get a well controlled magnetic field in a transverse cross section is to distribute the magnetic material all around the atomic beam to make a so-called Halbach cylinder. In the context of atom physics, fields with a linear or quadratic dependence have been used to realize refractive atom-optical components.\cite{Meschede1996, Doyle2007} Here a highly uniform field is required. Following Ref.~\onlinecite{Halbach1980} let us consider a magnetized rim such that the magnetization $\bm{M}$ at an angle $\theta$ from the $y$-axis makes an angle $2\theta$ with respect to the same axis (Fig.~\ref{Fig:Halbach}~(a)). Then, the magnetic field reads:
$$
\bm{B}_{\mathrm{Hal}}(\bm{r})= \left\{ \begin{array}{ll}
                                        \bm{0} \hspace{1cm} & \mathrm{for}\hspace{0.5cm}r>R_{\mathrm{ext}},\nonumber\\
                                        B_{\mathrm{R}}\ln\left(\displaystyle{\frac{R_{\mathrm{ext}}}{R_{\mathrm{int}}}}\right)\hat{\bm{y}}\hspace{1cm} &  \mathrm{for}\hspace{0.5cm}r<R_{\mathrm{int}},\nonumber
                                        \end{array}\nonumber
                                \right.\nonumber
$$
where $B_{\mathrm{R}}$ is the remanent field of the magnetic material, commonly in the $10-15~\mathrm{kG}$ range for modern rare-earth magnets.
Numerical investigations (see next section) indicate that a $8$-pole Halbach-like configuration as depicted in Fig.~\ref{Fig:Halbach}~(b) is able to produce fields on the order of $600~\mathrm{G}$ with homogeneity better than $1~\mathrm{G}$ on a $16~\mathrm{mm}$ cross section. Higher field strength and/or beam diameters are easy to achieve if necessary.

More detailed studies demonstrate that deviations on a typical $600~\mathrm{G}$ magnetic field stay below the $\pm1~\mathrm{G}$ limit for $\pm0.2~\mathrm{mm}$ mispositioning of the magnets, which is a common requirement on machining. Likewise, the same variations are observed for $\pm2.5\%$ dispersion in the strength of the magnets. This value is consistent with a rough statistical analysis we made on a sample of $25$ magnets.

\section{\label{sec:Field:calculations} Field calculations}
    \subsection{\label{subsec:Field:calculations:Magnets} Magnets modeling}
        \subsubsection{\label{subsubsec:Field:calculations:Magnets} Magnetic material}
Our setup uses long $2a\times2b\times2c=6\times6\times148~\mathrm{mm^3}$ NdFeB magnets (HKCM, part number: Q148x06x06Zn-30SH). They are made from $30\mathrm{SH}$ grade which has a higher maximum operation temperature than other grades. Its remanent field $B_\mathrm{R}=10.8~\mathrm{kG}$ is also lower. The device is thus more compact and outer field extension is reduced. Such rare-earth material is very hard from a magnetic point of view so very little demagnetization occurs when placed in the field of other magnets, at least in our case where fields do not exceed the kiloGauss range. This makes field calculations particularly simple and reliable. Even if an exact formula for the field of a cuboid magnet can be found,\cite{SuppInfo} in many cases, it can be replaced with an easy to handle dipole approximation.

        \subsubsection{\label{subsubsec:Field:calculations:Dipole} Dipole approximation}
In the proposed geometry described below, the magnets have a square cross section ($2b=2a$) and the long magnets can be decomposed in a set of cubic magnets with side $2a$. Then, one easily checks numerically that when the distance to the magnet is larger that twice the side, the field of the associated dipole is an accurate approximation of that of the actual magnet to better than $2\%$.\footnote{convergence can be much slower for cuboids with different aspect ratios.} It is not a very restrictive condition as in our case, $2a=6~\mathrm{mm}$ and magnets cannot be located nearer than $11~\mathrm{mm}$ from the beam axis.

A full vector expression of the field of a dipole can be found in any textbook. It is well adapted for computer implementation.
Even if the full magnetic system is then represented by more than $1500$ dipoles, calculations are still very fast: the simulations presented next section take less than one second on a conventional personal computer.\cite{SuppInfo}

    \subsection{\label{subsec:Field:calculations:Magnets:layout} Magnets layout}
\begin{figure}
\includegraphics[width=\columnwidth]{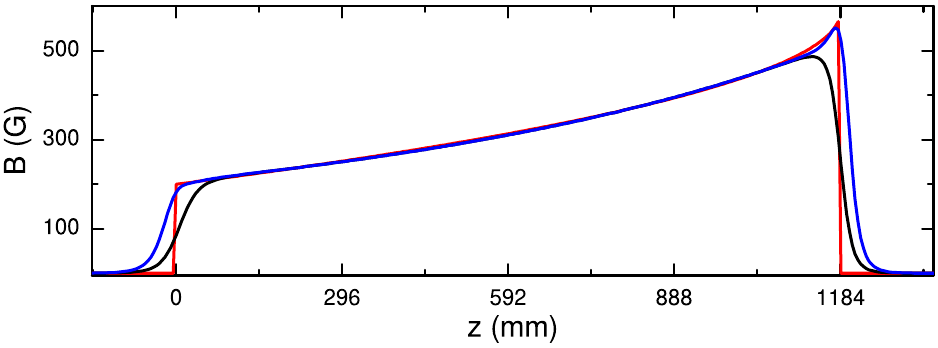}
\caption{Color online. Ideal (red) and calculated profiles, without (black) and with (blue) end caps.\label{Fig:CalcHalbach}}
\end{figure}
In principle, the field magnitude can be adjusted varying the amount, the density and/or the position of the magnetic material. The availability of very elongated magnets ($c/a\approx25$) directed us toward a simple layout. Only the distance to the axis $d(z)$ is varied. At first approximation the magnets can be considered as infinite. The magnetic field strength then decreases as the inverse of the distance squared. So, to produce the field $B(z)$ a good ansatz for $d(z)$ is:
\begin{equation}
\label{Eqn:DistanceToAxis}
d(z)=d(L)\sqrt{\frac{B(L)}{B(z)}}.
\end{equation}

As a matter of fact, this guess turns out to be both very efficient and close to a linear function. Numerical calculations show (Fig.~\ref{Fig:CalcHalbach}) that a linear approximation of Eq.~(\ref{Eqn:DistanceToAxis}) can be optimized to give a field within $\pm3~\mathrm{G}$ from the ideal one over the most part of the slower. Such deviations are completely irrelevant concerning the longitudinal motion. Magnets are then positioned on the generatrices of a cone and the mechanics is straightforward (Sec.~\ref{subsec:Mechanics:and:field:characterization:Mechanical:design}).

Naturally, the agreement is not so good at both ends where the ideal profile has sharp edges, while the actual field spreads out and vanishes on distances comparable to the diameter on which magnets are distributed. The actual $\Delta B$ is reduced which lowers the capture velocity and thus the beam flux. We made additional sections of eight extra cubic magnets in a Halbach configuration designed to provide localized improvement on the field profile at both ends (`{\em end caps}'). As seen on Fig.~\ref{Fig:CalcHalbach}, matching to the ideal profile is enhanced, especially at the high field side where the ideal profile exhibits a marked increase.

The length of the Zeeman slower is $L=1184~\mathrm{mm}$ corresponding to eight sections of $148~\mathrm{mm}$-long magnets. The capture velocity is then $v(0)=(2\eta a_{\mathrm{max}}L)^{1/2}\approx450~\mathrm{m\cdot s^{-1}}$ and $\Delta B=388~\mathrm{G}$. A bias field $B_{\mathrm{bias}}=200~\mathrm{G}$ is added to avoid low-field level crossings around $120~\mathrm{G}$. These field parameters together with the magnet size and properties determine the distance and angle from axis of the magnets. In our case, the best choice was a slope of $-15.9~\mathrm{mm/m}$ corresponding to $d(0)=49.5~\mathrm{mm}$ and $d(L)=30.7~\mathrm{mm}$. Entrance and output end caps are both made of $10~\mathrm{mm}$-side cubic magnets of N35 grade ($B_{\mathrm{R}}=11.7~\mathrm{kG}$). They are located on circles whose diameters are $94.0~\mathrm{mm}$ and $66.0~\mathrm{mm}$ respectively.

\section{\label{sec:Mechanics:and:measurements} Mechanics and field measurements}
    \subsection{\label{subsec:Mechanics:and:field:characterization:Mechanical:design} Mechanical design}
\begin{figure}
\includegraphics[width=\columnwidth]{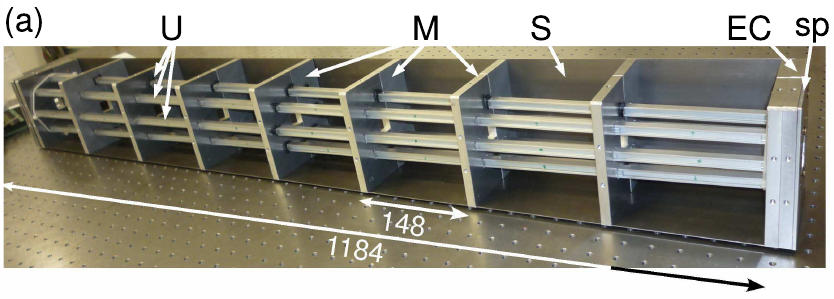}
\includegraphics[width=\columnwidth]{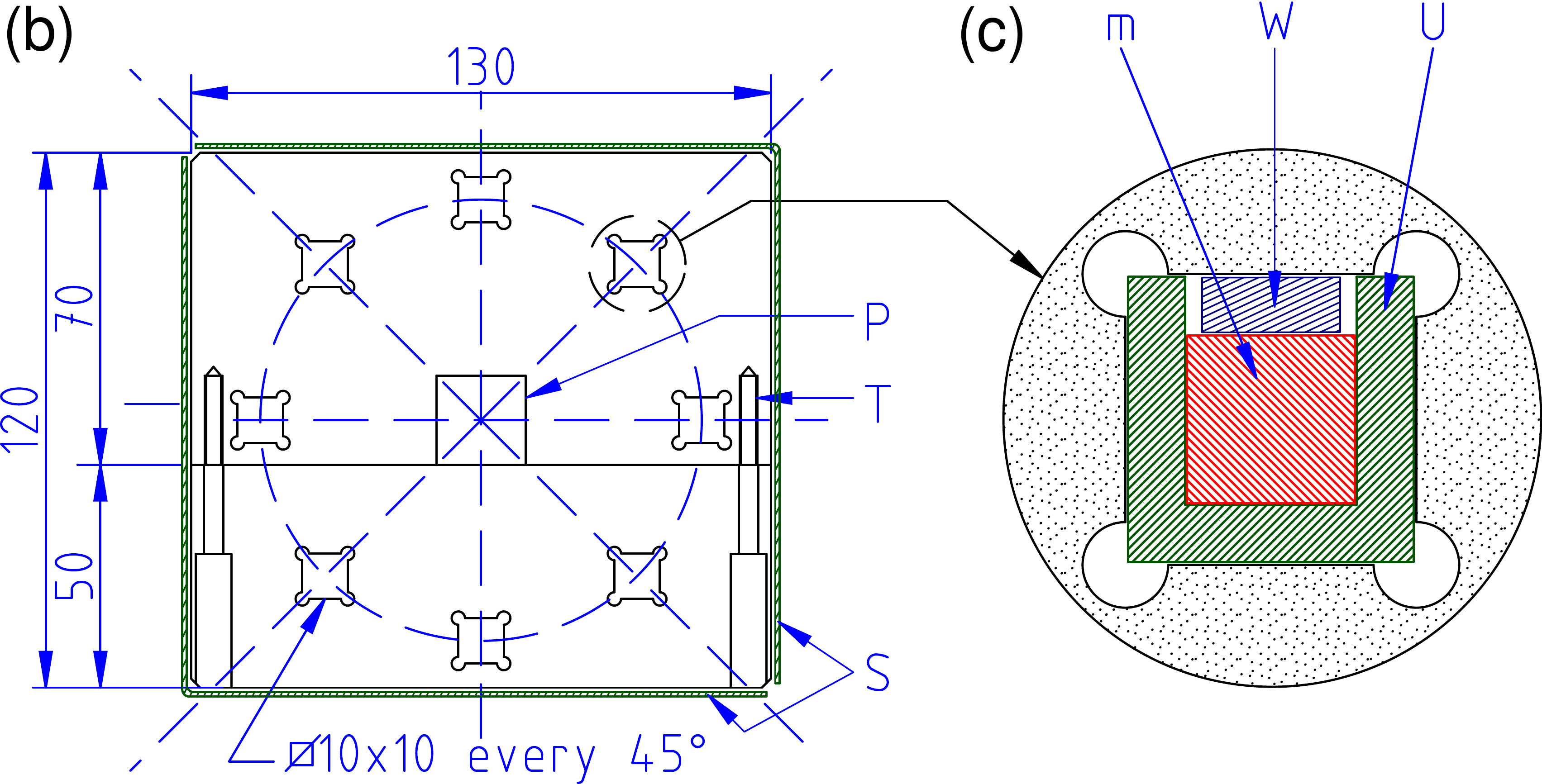}
\caption{Color online. (a) Picture of the Zeeman slower: [M] mounts, [EC] output end cap screwed in last mount, [U] U-shaped profiles, [S] half part of the shield, [sp] $5~\mathrm{mm}$ spacer between end cap and shield side. (b) Individual mount; [T] threading to screw the two parts of the mount together, [P] central square milling in which CF16 pipe goes through. (c) detail of a square hole to show U-shaped profiles insertion, magnets [m] and plastic wedge [W]. Dimensions in mm.\label{Fig:techdraw}}
\end{figure}

The Zeeman slower consists in 9 mounts supporting 8 U-shaped aluminum profiles (Fig.~\ref{Fig:techdraw}). The U-shaped profiles go through the mounts by means of square holes evenly spaced on a circle whose diameter decreases from mount to mount according to Eq.~(\ref{Eqn:bid}) and Eq.~(\ref{Eqn:DistanceToAxis}) possibly linearized. Magnets are then inserted one after each other in the U-shaped profiles and clamped by a small plastic wedge. End caps are filled with the suitable block magnets and screwed together with their spacer in the first and last mount (see Fig.~\ref{Fig:techdraw}~(a)). The whole setup is then rigid and all parts tightly positioned. Indeed, as said before, calculations are very reliable and Zeeman slower operation is known to be robust so there is no need for adjustment. Mounts are made of two parts screwed together. The Zeeman slower can then be assembled around the CF16 pipe without vacuum breaking {\em e.g.} after baking out the UHV setup.\cite{SuppInfo}

        \subsection{\label{subsec:Mechanics:and:field:characterization:Shielding} Shielding}
Stray magnetic fields might strongly affect atomic physics experiments. Actually, the 8-pole configuration produces very little field outside (see Fig.~\ref{Fig:B_trans}~(a)), except of course, at both ends. However, to lower stray fields even further, we have made a rectangular single-layer shield from a $1~\mathrm{mm}$-thick soft iron sheet wrapped around the mounts. Besides, mechanical properties and protection are also improved. As seen on Fig.~\ref{Fig:B_longit}~(a), the inner field is almost unaffected. On the contrary, the outer magnetic field falls down much quicker all the more since the plateau around $0.5~\mathrm{G}$ in Fig.~\ref{Fig:B_longit}~(b) is probably an artifact associated with the probe. In practice, no disturbance is detected on the MOT and even on optical molasses $125~\mathrm{mm}$ downstream.\cite{SuppInfo}

    \subsection{\label{subsec:Mechanics:and:field:characterization} Magnetic field and lasers characterization}
        \subsubsection{\label{subsubsec:Mechanics:and:field:characterization:field:parameters} Magnetic field}

\begin{figure}
\includegraphics[width=\columnwidth]{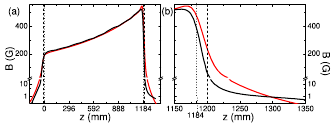}
\caption{Color online. Calculated (red) and measured (black) magnetic field profiles. (a) Scan along the beam axis. (b) Close up of the output region. In the calculation the shield is not taken into account. Dotted and dashed lines indicate the Zeeman slower and the shield physical ends. Log scale before break.\label{Fig:B_longit}}
\end{figure}

Magnetic field measurements are done with a home-made 3D probe using 3 Honeywell SS495 Hall effect sensors.\cite{SuppInfo} Figure~\ref{Fig:B_longit} displays a longitudinal scan of the magnetic field on the axis of the Zeeman slower with end caps and shield. It can be first noticed that the longitudinal profile is intrinsically very smooth as the magnets make a uniform magnetized medium throughout the Zeeman slower. After calibration of the magnetic material actual remanent field, deviations from the calculated profile are less than a few Gauss. Besides, one usually observes only localized mismatches attributed to the dispersion in the strength of the magnets. The shield input and output sides flatten the inner field at both ends. Of course the effect decreases when they get further apart but the Zeeman slower should not be lengthened too much. A $5~\mathrm{mm}$ spacer (tag [sp] in Fig.~\ref{Fig:techdraw}) is a good trade off. Then, the actual magnetic field measured parameters are $B_{\mathrm{bias}}=200~\mathrm{G}$ and $\Delta B=350~\mathrm{G}$ only slightly smaller than the calculated value.

\begin{figure}
\includegraphics[width=\columnwidth]{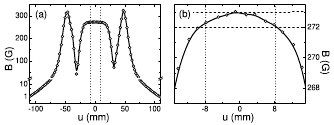}
\caption{(a) Measured magnetic field without shield across the beam axis at $z\sim460~\mathrm{mm}$ along the $u$-direction of Fig.~\ref{Fig:Halbach}. (b) Close up of the central region. Dashed lines indicate the atom beam extension and a $1~\mathrm{G}$ magnetic field span. Line to guide the eye. Log scale before break. The shield was removed to allow the probe to go through. With the shield, the inner field is almost unaffected an the outer field is below the probe sensitivity.\label{Fig:B_trans}}
\end{figure}

Figure~\ref{Fig:B_trans} depicts a transverse cut of the magnetic field. It is realized along the $u$-direction of Fig.~\ref{Fig:Halbach} near the middle of the Zeeman slower ($z\sim460~\mathrm{mm}$). The shield was removed to allow the probe to go through. It exhibits the two expected features: (i) little outer field (ii) highly homogeneous inner field. In the vicinity of the axis, the measured profile is however less flat than expected. This is mainly due to the finite size of the probe. Anyway, magnetic field deviations stay within a Gauss or so in the region of interest. With the shield, the outer field is below probe sensitivity.

\subsubsection{\label{subsubsec:Mechanics:and:field:characterization:field:lasers} Lasers}

The Zeeman slower operates between the $5^2S_{1/2}$ and $5^2P_{3/2}$ states of $^{87}$Rb around $\lambda=780~\mathrm{nm}$ (D2 line). For an increasing-field Zeeman slower, a closed $\sigma^-$ transition is required,\cite{barrett1991} $F=2,\;m_F=-2\leftrightarrow F^\prime=3,\;m_{F^\prime}=-3$ in our case. However, the magnetic field is here perpendicular to the propagation axis. Thus, any incoming polarization state possesses {\em a priori} $\pi$ and $\sigma^\pm$ components: it is not possible to create a pure $\sigma^-$ polarization state (see Fig.~\ref{Fig:longitrans} and Ref. \onlinecite{Grynberg2010}). In addition to laser power losses, the $\pi$ and $\sigma^+$ components excite the $m_{F^\prime}=-2$ and $-1$ states from which spontaneous emission populates $F=1$ ground state levels. Repumping light is thus necessary between the $F=1$ and $F^\prime=2$ manifolds. The detrimental effect of the unwanted polarization components is minimized when the incoming polarization is perpendicular to the magnetic field since there is no $\pi$ contribution in that case. We measured a $20^\circ$ (FWHM) acceptance for the polarization alignement.

Permanent magnets enable to easily reach magnetic fields on the order of $B_{\mathrm{bias}}+\Delta B\approx500-600~\mathrm{G}$. As a consequence, detuning of the cycling light below the transition frequency amounts to $\delta_0\approx-800~\mathrm{MHz}$ (Eq.~(\ref{Eqn:delta0})). Such high detunings are realized sending a master laser through two double pass $200~\mathrm{MHz}$ AOMs before locking on a resonance line using saturation spectroscopy. The repumper is simply locked on the red-detuned side of the broad Doppler absorption profile.

The two master lasers are Sanyo DL7140-201S diodes having a small linewidth ($\sim5~\mathrm{MHz}$). We use them without external cavity feedback. Beams are recombined on a cube and pass through a polarizer. Then they are sent with the same polarization into a 1W Tapered Amplifier (Sacher TEC-400-0780-1000). A total power of more than $250~\mathrm{mW}$ is available on the atoms after fiber coupling. The beam is expanded to about $23~\mathrm{mm}$ (full width at $1/e^2$) and focused in the vicinity of the oven output aperture for better transverse collimation of the atomic beam.

\section{\label{sec:Experimental:apparatus} Experimental apparatus}
    \subsection{\label{subsec:Experimental:apparatus:Vacuum} Vacuum system}
\begin{figure}
\includegraphics[width=\columnwidth]{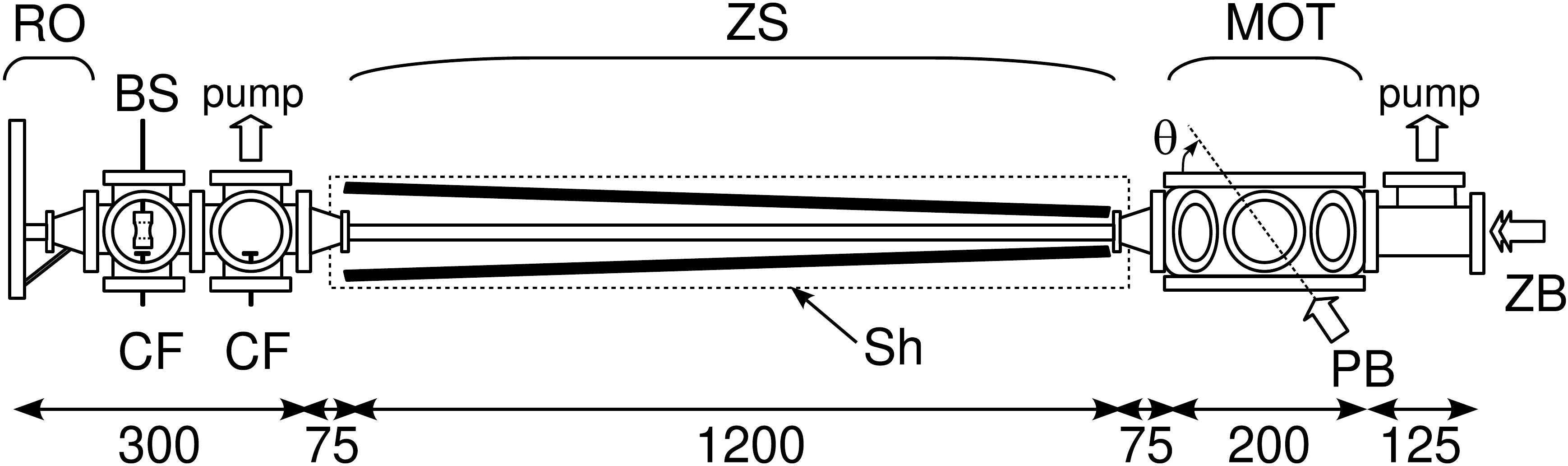}
\caption{Sketch of the overall experimental setup. [RO] recirculating oven, [BS] beam shutter, [CF] cold finger, [ZS] Zeeman slower, [MOT] MOT chamber, [ZB] Zeeman cycling and repumping beams, [PB] $\theta=56^\circ$ probe beam, [Sh] magnetic shield. $45^\circ$ and $90^\circ$ probe beams are sent through the horizontal windows depicted on the MOT chamber. Dimensions in mm, not rigorously to scale.\label{Fig:manip}}
\end{figure}
Figure~\ref{Fig:manip} shows a sketch of the experimental setup. At one end, the MOT chamber is a spherical octagon from Kimball physics (MCF600-SO200800). It has two horizontal CF100 windows and eight CF40 ports. It is pumped by a $20~\mathrm{L/s}$ ion pump. One CF40 port is connected to the $1200~\mathrm{mm}$-long CF16 pipe around which the Zeeman slower is set. At the other end, one finds a first 6-way cross, used to connect a $40~\mathrm{L/s}$ ion pump, a thermoelectrically-cooled cold finger and two windows for beam diagnosis. It is preceded by a second 6-way cross that holds another cold finger, a angle valve for initial evacuation of the chamber and a stepper-motor-actuated beam shutter. Finally, the in-line port holds the recirculating oven.\cite{SuppInfo}

    \subsection{\label{subsec:Experimental:apparatus:Probe} Probe beams}
Probe beams on the $F=2\leftrightarrow F^\prime=3$ transition can be sent in the chamber through the different windows and absorption is measured in this way at $45^\circ$, $56^\circ$ or $90^\circ$ from the atomic beam. Absorption signals are used to calibrate fluorescence collected through a CF40 port by a large aperture condenser lens and focused on a $1~\mathrm{cm}^2$ PIN photodiode (Centronics OSD 100-6). Photocurrent is measured with a homemade transimpedance amplifier (typically $10~\mathrm{M}\Omega$) and a low-noise amplifier (Stanford Research Systems SR560) used with a moderate gain ($G=5$) and a $3~\mathrm{kHz}$ low-pass filter. Frequency scans are recorded on a digital oscilloscope and averaged for 8-16 runs. During the measurements, a repumper beam on the $F=1\rightarrow F^\prime=2$ transition may be turned on.

\section{\label{sec:Zeeman:Slower:performances} Zeeman slower performances}
    \subsection{\label{subsec:Zeeman:Slower:performances:Atom:flux} Atom flux}
\begin{figure}
\includegraphics[width=\columnwidth]{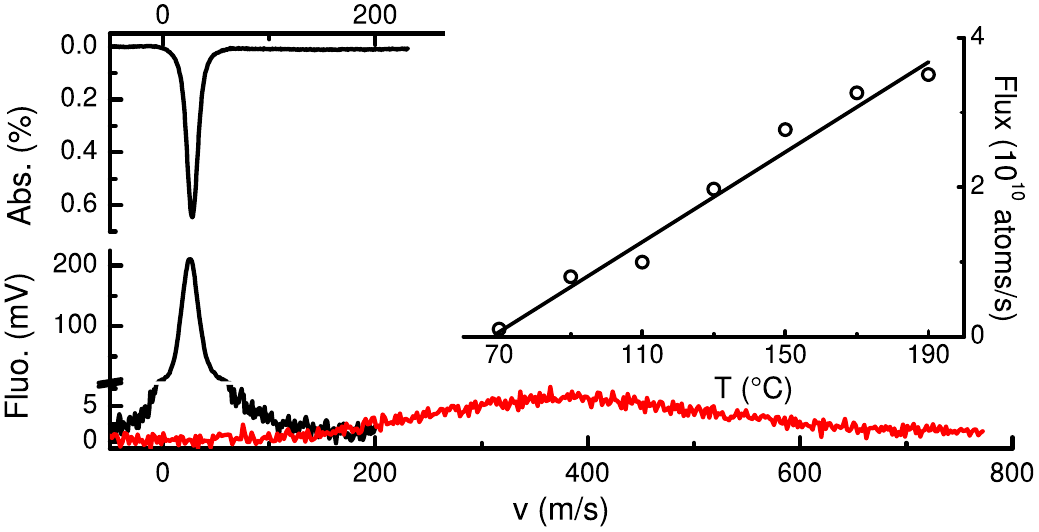}
\caption{Color online. Red: thermal beam fluorescence signal. Black: absorption and fluorescence signals of the slowed beam; axis break on fluorescence signal. Inset: temperature dependence of the atom flux; line to guide the eye.\label{Fig:fluo}}
\end{figure}
Figure~\ref{Fig:fluo} displays typical fluorescence and absorption signals. The oven base temperature was set to $T_1=190^\circ$C so that fluorescence of the thermal unslowed beam is clearly visible. When Zeeman light is on, a sharp peak at low velocity appears both in the fluorescence and absorption spectra. Detuning of the cycling light in that experiment is such that the final velocity is about $25~\mathrm{m\cdot s^{-1}}$.

These signals are recorded scanning the frequency of a probe beam making an angle $\theta$ with the atomic beam. A given detuning $\Delta$ of the probe from resonance corresponds to the excitation of the velocity class $v=\lambda\Delta/\cos\,\theta$. The absorption signal $A(\Delta)$ is then converted into $A(v)$ as in Fig.~\ref{Fig:fluo} from which typical output velocity $\overline{v}$, velocity spread $\delta v$ and maximum absorption $A_\mathrm{max}$ can be estimated. The atom flux $\Phi$ then reads:
$$
\label{Eqn:atomflux}
\Phi=c\sin\theta\cos\theta D\frac{A_\mathrm{max}\overline{v}\delta v}{\lambda \Gamma \sigma_0},
$$
where $c$ is a numerical parameter near unity;\cite{SuppInfo} $\Gamma$, $\sigma_0$ and $D$ denote the transition decay rate, the resonant cross section and atomic beam diameter.

On a separate experiment, we spatially scan a small probe beam across the atomic beam. The atom density exhibits a trapezoid shape. The measured length of the parallel sides are $20$ and $30~\mathrm{mm}$ so we take $D=25~\mathrm{mm}$. It corresponds well to the free expansion of the collimated beam from the CF16 output of the Zeeman slower. Then, the typical estimated flux for a maximum absorption $A_\mathrm{max}=0.6\%$ is $\Phi=4\times10^{10}~\mathrm{atoms/s}$.

The flux increase with oven temperature is plotted in the inset of Fig.~\ref{Fig:fluo}. Typical experiments are carried at $T_1=130^\circ$C for which we get an intense slow beam of $2\times10^{10}~\mathrm{atoms/s}$.

Finally, we measured little influence of the entrance end cap on the atom flux and a moderate increase, $10\pm5\%$, with the output one.
    \subsection{\label{subsec:Zeeman:Slower:performances:Velocity:distribution} Velocity distribution}
\begin{figure}
\includegraphics[width=\columnwidth]{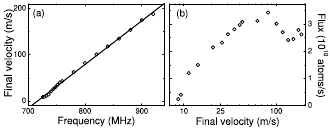}
\caption{(a) Final velocity as a function of Zeeman cycling light detuning ($I=4.7~\mathrm{mW\cdot cm^{-2}}$). Line: linear fit, the slope is $0.95~\mathrm{m\cdot s^{-1}/MHz}$. (b) Atom flux as a function of final velocity.\label{Fig:speed_freq_flux}}
\end{figure}
Naturally, the Zeeman cycling light detuning strongly affects the atom beam velocity distribution (Fig.~\ref{Fig:speed_freq_flux}).
A linear dependence of the final velocity in the detuning is observed. The actual slope is on the order of that expected from a simple model $\mathrm{d}v/\mathrm{d}\delta=2\pi/k=0.78~\mathrm{m\cdot s^{-1}/MHz}$ but slightly higher and intensity-dependent.\cite{bagnato1991}

Besides, the atom flux is roughly constant for final velocities above $40~\mathrm{m\cdot s^{-1}}$. Below this value, the flux measured in the chamber $125~\mathrm{mm}$ downstream decreases. Indeed, the beam gets more divergent and atoms are lost in collisions with the walls of the vacuum chamber.
    \subsection{\label{subsec:Zeeman:Slower:performances:Necessary:powers} Needed laser powers}
\begin{figure}
\includegraphics[width=\columnwidth]{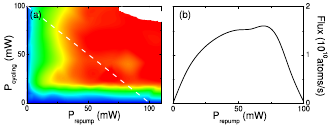}
\caption{Color online. (a) Atom flux as function of cycling and repumper beams powers. (b) Cross section along the white dotted line corresponding to a total available power of $100~\mathrm{mW}$. Power ratio is measured with a scanning Fabry-Perot interferometer. \label{Fig:P_rep_cycl}}
\end{figure}
Figure~\ref{Fig:P_rep_cycl} demonstrates that comparable amounts of cycling and repumper light are necessary. With a total power of $100~\mathrm{mW}$ we get a non-critical operation of the Zeeman slower at its best flux and a final velocity of $30~\mathrm{m\cdot s^{-1}}$, well suited for efficient MOT loading. The equivalent intensity is about $24~\mathrm{mW\cdot cm^{-2}}$. However, as we shall see now, a lot of power can be saved with a more elaborate strategy.

    \subsection{\label{subsec:Zeeman:Slower:performances:Repumper} Repumper}
\begin{figure}
\includegraphics[width=\columnwidth]{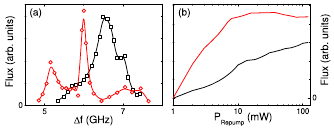}
\caption{Color online. (a) Atom flux as a function of repumper frequency. $\Delta f$ is the beat note frequency of the repumper with an auxiliary laser locked on the $F=2\rightarrow F^\prime=3$ resonance line; red circles/black squares: repumper polarization perpendicular/parallel to the magnetic field. (b) Atom flux as a function of repumper power (log scale) when its frequency is fixed (black) or swept (red) across the full spectrum of left panel.\label{Fig:repump}}
\end{figure}
In the results reported until now, repumping and cycling light have the same polarization: linear and perpendicular to the magnetic field, a state commonly referred to as {\em linear $\sigma$} recalling that it is a superposition of $\sigma^\pm$ states.\cite{Grynberg2010} If no common amplification in a tapered amplifier is used, polarizations are likely to be orthogonal. The repumper polarization is then parallel to the magnetic field {\em i. e.} a $\pi$ state. When the repumper frequency is varied as in Fig.~\ref{Fig:repump}~(a) very different spectra for the two configurations are observed. Efficient repumping occurs with more or less well defined peaks spread over about $2~\mathrm{GHz}$ and roughly centered around the $F=1\to F'=2$ transition. This means that several depumping/repumping pathways are involved, probably occurring at localized places along the Zeeman slower.

It is not easy to get a simple picture of what is happening: a complete \emph{ab-initio} simulation of the internal dynamics is not simple due to the large number of Zeeman sublevels (24 in total for all the ground and excited states), the multiple level crossings occurring in the 50--200~G range, and high light intensities. However, one can overcome this intricate internal dynamics by sweeping quickly (typically around 8~kHz) the repumper frequency over all the observed peaks. With a low-pass filter, the central frequency remains locked on the side of the Doppler profile. Doing so, we get a slightly higher flux for significantly less repumper power, typically 10 mW (Fig.~\ref{Fig:repump}~(b)).

    \subsection{\label{subsec:Zeeman:Slower:performances:MOT:loading} MOT loading}
A final demonstration of the Zeeman slower efficiency is given by monitoring the loading of a MOT. It is made from 3 retroreflected beams $28~\mathrm{mm}$ in diameter (FW at $1/e^2$). We use $10-20~\mathrm{mW}$ and $1-3~\mathrm{mW}$ of cycling and repumper light per beam. When the Zeeman slower is on with a final velocity of $30~\mathrm{m\cdot s^{-1}}$, a quasi exponential loading is observed with characteristic time $\tau\sim320~\mathrm{ms}$ for magnetic field gradients on the order of $15-20~\mathrm{G\cdot cm^{-1}}$. After one second or so, the cloud growth is complete. From absorption spectroscopy, we deduce a density $n=1.4\times10^{10}~\mathrm{atoms\cdot cm^{-3}}$. The typical cloud size is $12~\mathrm{mm}$ so we estimate the atom number to be on the order of $N=2\times10^{10}$. These figures are consistent with the above measurements of an atom flux of several $10^{10}~\mathrm{atoms/s}$ and nearly unity capture efficiency. As expected, thanks to the high magnetic field in the slower, the Zeeman beams are far detuned and do not disturb the MOT.

\section{\label{sec:Conclusion} Conclusion}
We have presented a simple and fast to build, robust Zeeman slower based on permanent magnets in a Halbach configuration. Detailed characterization shows it is an efficient and reliable source for loading a MOT with more than $10^{10}$ atoms in one second. Without power nor cooling water consumption, the apparatus produces homogeneous and smooth high fields over the whole beam diameter and low stray fields. It also simplifies high-temperature bakeout. We thus believe it to be a very attractive alternative to wire-wound systems.

\begin{acknowledgments}
We thank J. M. Vogels for his major contribution to the design of the recirculation oven and D. Comparat for useful bibliography indications.
This work was supported by the Agence Nationale pour la Recherche (ANR-09-BLAN-0134-01), the R\'egion Midi-Pyr\'en\'ees, and the Institut Universitaire de France.
\end{acknowledgments}

\onecolumngrid
\newpage
\vskip 1cm
\noindent
\begin{center}
{\large \bf Supplementary material for \\ Zeeman slowers made simple with permanent magnets in a Halbach configuration}
\end{center}
\vskip 1cm


\setcounter{section}{0}
\setcounter{figure}{0}
\setcounter{page}{1}

\renewcommand\thepage{S\arabic{page}}
\renewcommand\thesection{S\arabic{section}}
\renewcommand\thefigure{S\arabic{figure}}

\section{\label{sec:Introduction} Introduction}

We collect here some extra information that may be useful to the reader of the main paper.

\begin{itemize}
  \item Section~\ref{Magnetic:field:for:a:cuboid} gives exact formulas for the magnetic field of a cuboid {\em e. g.} if the dipole decomposition described in the paper is to be avoided.
  \item Section~\ref{3D:probe} is a more detailed description of the 3D-probe we use for field measurements (see Figs.~\ref{Fig:planche1} (d) and (e)).
  \item In section~\ref{Flux} we give the formula used to compute the atom flux from absorption measurements.
  \item Section~\ref{Oven} is a description of our recirculating oven whose design is unpublished.
  \item Finally, we present some pictures of the building of our Zeeman slower in section~\ref{Pictures}.
\end{itemize}

\section{\label{Magnetic:field:for:a:cuboid} Magnetic field for a cuboid}
\begin{figure}[!h]
\center
\includegraphics[width=0.6\columnwidth]{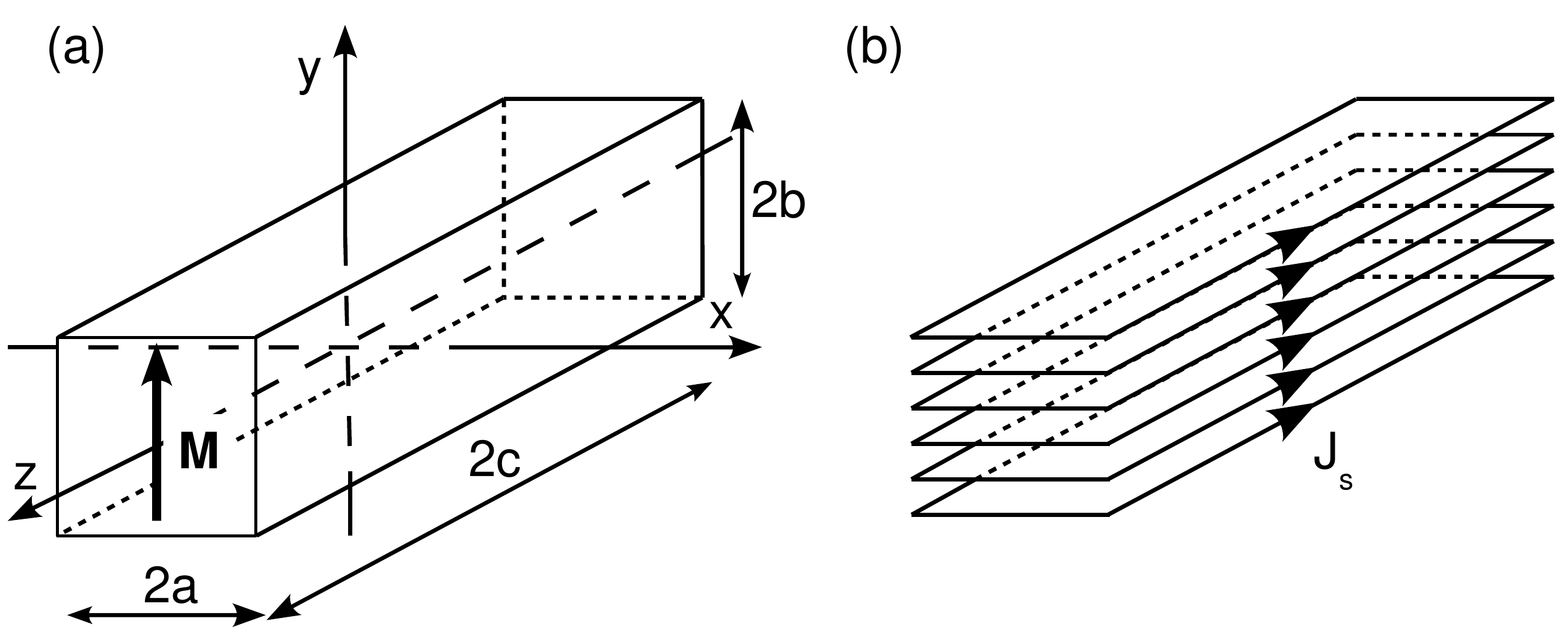}
\caption{Left: conventions and notations for a block magnet. Right: equivalent rectangular solenoid.\label{Fig:axes}}
\end{figure}
Let us consider a $2a\times2b\times2c$ cuboid magnet with magnetization $\mathbf{M}$ along the $y$-axis (see Fig.~\ref{Fig:axes}(a)).
As the magnetic field for a segment is analytically known, that for a rectangular coil is easy to calculate. Introducing two auxiliary functions:$$F(X,Z)=\frac{XZ}{\left(X^2+y^2\right)\sqrt{X^2+Y^2+Z^2}}\,\,\,\,\,\,\mathrm{and}\,\,\,\,\,\,G(X)=-\frac{y}{X},$$
the magnetic field for a current $I$ is $\bm{B}_{\mathrm{coil}}=\mu_0I\bm
{b}$ with:
\begin{eqnarray}
\label{Eqn:coilb}
b_x(x,y,z)&=&\sum_{p,q=0}^{1}(-1)^{p+q}F(x-{\left(-1\right)}^pa,z-{\left(-1\right)}^qc)G(x-{{\left(-1\right)}^p}a),\nonumber\\
b_y(x,y,z)&=&\sum_{p,q=0}^{1}(-1)^{p+q}
\left(  F(x-{\left(-1\right)}^pa,z-{\left(-1\right)}^qc)+
        F(z-{\left(-1\right)}^pc,x-{\left(-1\right)}^qa)
\right),\nonumber\\
b_z(x,y,z)&=&\sum_{p,q=0}^{1}(-1)^{p+q}F(z-{\left(-1\right)}^pc,x-{\left(-1\right)}^qa)G(z-{{\left(-1\right)}^p}c).\nonumber
\end{eqnarray}
Integration along the $z$-axis can then be computed and the field for a rectangular solenoid reads:
\begin{equation}
\label{Eqn:solenoidB}
\bm{B}_{\mathrm{sol}}(x,y,z)=\mu_0J_s\sum_{n,p,q=0}^{1}(-1)^{n+p+q}\bm{\mathcal{B}}\left(x-(-1)^pa,y-(-1)^nb,z-(-1)^qc\right)
\end{equation}
where $\bm{\mathcal{B}}$ is a vector field whose coordinates are:
\begin{eqnarray}
\label{Eqn:solenoidBeta}
\mathcal{B}_x(X,Y,Z)&=&{\frac{1}{2}}\log \left({\frac{{\sqrt{{X^2} + {Y^2} + {Z^2}}}-Z}{{\sqrt{{X^2} + {Y^2} + {Z^2}}}+Z}}\right)\nonumber \\
\mathcal{B}_y(X,Y,Z)&=&-\arctan\left({\frac{Y\,{\sqrt{{X^2} + {Y^2} + {Z^2}}}}{X\,Z}}\right)\nonumber\\
\mathcal{B}_z(X,Y,Z)&=&\mathcal{B}_x(Z,Y,X)\nonumber.
\end{eqnarray}
In Eq.\ref{Eqn:solenoidB}, $J_s$ is the equivalent surface current density whose magnitude identifies with the magnetization $M$ for the considered cuboid (Fig.~\ref{Fig:axes}). NdFeB is such a hard magnetic material that, in our case, $\mu_0J_s=\mu_0M$ can safely be taken equal to $B_\mathrm{R}$ to get the desired magnetic field.

\section{\label{3D:probe} 3D probe}
\begin{figure}[!h]
\begin{center}
\includegraphics[width=16cm]{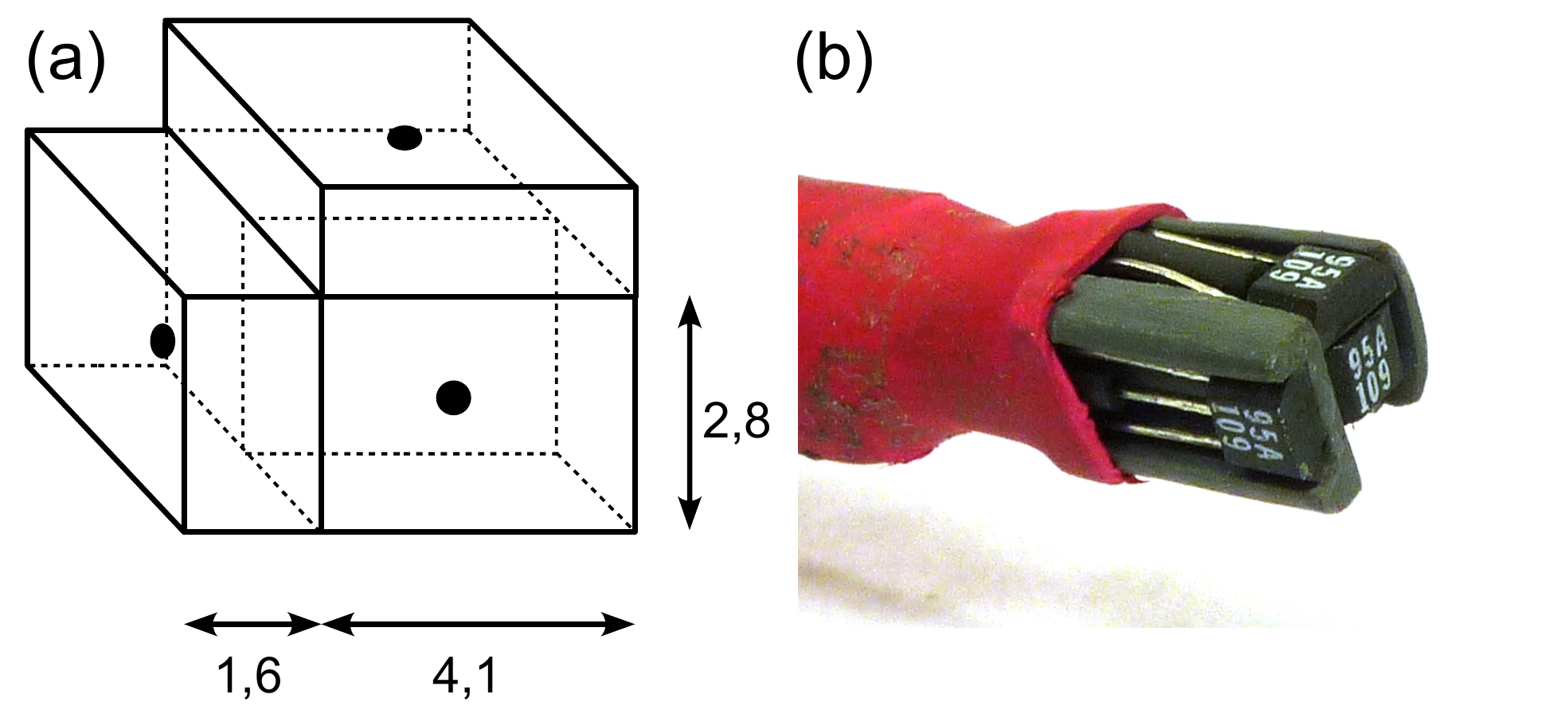}
\caption{(a) Sketch of our 3D-probe. Black disks represent sensing area. Approximate dimensions in mm. Not rigorously to scale. (b) Picture of our probe.}
\label{Fig:probe}
\end{center}
\end{figure}

A 3D magnetic probe was constructed using 3 Honeywell SS495 Hall effect sensors connected to a Keithley datalogger. It was calibrated with a commercial 1D Leybold probe. Accuracy is estimated to be $3\%$. Measurements below the $1~\mathrm{G}$ level should be considered carefully regarding the low sensitivity of the probe and difficulties in background subtraction. Besides, the three components of the magnetic field are measured at different locations (typically $5~\mathrm{mm}$) apart from each other due to sensors physical dimensions (see Fig.~\ref{Fig:probe}). This is the cause of some inaccuracy related to field gradients.

The probe is guided in a plastic pipe passing through the Zeeman slower. It is moreover attached to a flexible wire that winds around the axis of a stepper motor. When operated at a given frequency, the probe is smoothly translated and the magnetic field is recorded with a typical $2-3~\mathrm{points/mm}$ spatial resolution.

\section{\label{Flux} Atom flux calculation}
Assuming cylindrical symmetry, let $n(\rho)$ denote the beam density at a distance $\rho$ from the center and $f(v)$ the atomic velocity distribution. The atom flux is:
\begin{equation}
\label{Eqn:PhiDelta}
\Phi=\int_{0}^{+\infty}\!\!\!\!\!\! 2\pi\rho\;\mathrm{d}\rho\; n(\rho)\int_{0}^{+\infty}\!\!\!\!\!\! \mathrm{d}v\; v f(v).
\end{equation}
For a probe beam at an angle $\theta$ from the atomic beam, the absorption signal $A(\Delta)$ is:
\begin{equation}
\label{Eqn:ADelta}
A(\Delta)=\int_{-\infty}^{+\infty}\!\! \frac{\mathrm{d}\rho}{\sin\theta}\;\; n(\rho)\int_{0}^{+\infty}\!\!\!\!\!\! \mathrm{d}v\; f(v)\sigma(v,\Delta),
\end{equation}
where $\sigma(v,\Delta)=\sigma_0/(1+4\Delta^{\prime2}/\Gamma^2)$ with $\Delta^\prime=\Delta+kv\cos\theta$ is the absorption cross section. In the limit where the Doppler broadening is much larger than the natural linewidth, $k\,\delta v\cos\theta\gg\Gamma$, we can approximate the absorption cross section by a $\delta$-function: $$\sigma(v,\Delta)\approx\frac{\pi}{2}\sigma_0\delta\left(\frac{\Delta}{\Gamma}-\frac{v\cos\theta}{\lambda\Gamma} \right).$$
In that case Eq.~(\ref{Eqn:ADelta}) simplifies:
$$A(\Delta)=\frac{\pi}{2}\sigma_0\frac{\lambda\Gamma}{\cos\theta}f\left(\frac{\lambda\Delta}{\cos\theta}\right)\int_{-\infty}^{+\infty}\!\!\frac{\mathrm{d}\rho}{\sin\theta}\; n(\rho)$$
and Eq.~(\ref{Eqn:PhiDelta}) gives:
\begin{equation}
\label{Eqn:Flux}
\Phi=C\,D\frac{\lambda\tan\theta}{\sigma_0\Gamma}\int_{-\infty}^{+\infty}\!\!\!\!\!\! \mathrm{d}\Delta\;\Delta A(\Delta),
\end{equation}
where $D$ is the typical atom beam diameter and $C$ a constant near unity defined according to:
$$C=\frac{2}{\pi}\frac{1}{D}\frac{\int_{0}^{+\infty} 2\pi\rho\;\mathrm{d}\rho\; n(\rho)}{\int_{-\infty}^{+\infty} \mathrm{d}\rho\; n(\rho)}.$$
For a homogeneous cylindrical beam $C=1/2$. The atom flux is thus measured from the absorption signal by numerical integration following Eq.~(\ref{Eqn:Flux}).

\section{\label{Oven} Recirculating oven}
The atomic beam is created by an effusive oven loaded with 15~g of rubidium. In order to maximise the oven lifetime, we use a recirculating design. As compared to the so-called `candlestick' designs,\cite{hau1994,walkiewicz2000} our oven, inspired in part by Ref.~\onlinecite{carter1978} is very simple and easy to operate. We have built several versions of the oven over the last five years, with minor variations between them, and observed comparable performances. The same design has been used also for a sodium BEC setup producing extremely large condensates\cite{vanderstam2007} but no detailed description is given there.
\begin{figure}[!h]
\begin{center}
\includegraphics[width=8.6cm]{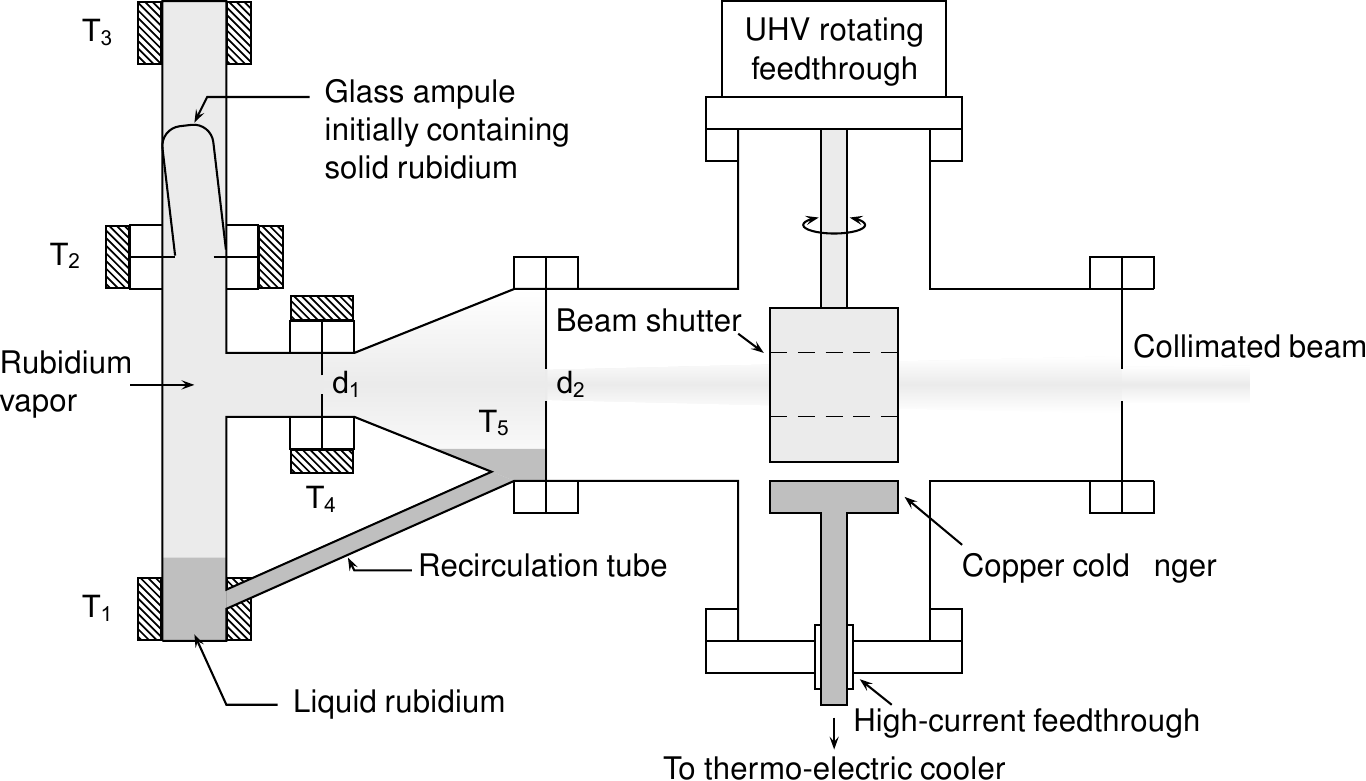}
\caption{Sketch of the recirculating oven. Hatched rectangles represent heaters.}
\label{Fig:oven}
\end{center}
\end{figure}
A general view of the oven, made of standard CF-16 and CF-40 ultra-high vacuum fittings, is shown on Fig.~\ref{Fig:oven}. A first chamber contains, at the bottom, molten rubidium kept at temperature $T_1$, in equilibrium with rubidium vapor, which effuses through a circular aperture of diameter $d_1=8$~mm drilled in the center of a blank CF-16 copper gasket. The other parts of the chamber are kept at $T_2=T_3=T_4\sim T_1+30$~K in order to avoid the accumulation of rubidium on a cold spot and possible clogging of the oven aperture.
The temperatures $T_1$ to $T_4$ are actively stabilized by means of four PID controllers, thermocouples as temperature sensors, and heating bands as actuators. To achieve a good thermal insulation, the oven is covered with two layers of alkaline earth silicate wool and an external foil of metal coated Mylar. In steady state, the average power consumption is a few tens of watts. A second chamber (made of a conical CF-16 to CF-40 adapter) is used to collimate the beam by means of a second aperture (diameter $d_2=4$~mm) located 80 mm downstream.

The rubidium not used in the collimated beam accumulates into this chamber. Liquid rubidium at temperature $T_5<T_1$ flows back by gravity to the first chamber through a 6~mm inner diameter stainless steel tube. A piece of gold-plated stainless steel mesh (Alfa-Aeser ref. 42011) covers the inside of the recirculating chamber to ease the accumulation of rubidium in its lower part.\footnote{in some versions of the oven we rolled a small quantity of this mesh into a `wick' that was inserted into the recirculation tube, in order to help recirculation by capillary action. However the ovens without this wick showed similar performance and the presence of the mesh is probably not necessary at all.}

The loading of the oven is made in a very simple way: we cool down the rubidium ampule(s) in liquid nitrogen, break the glass with pliers, and insert the ampules upside down into the first oven chamber. We then close this chamber with a blank CF-16 flange, and pump down the oven to $\sim 10^{-6}$~mbar with a turbomolecular pump. When heating the oven, the rubidium melts and drips to the bottom of the first chamber.

It is not easy to obtain a definitive proof that the rubidium does recirculate in the oven. However we have been running two ovens for several years, either at moderate or high flux.\cite{lahaye2005} In both cases, we did not observe any decrease in the atom flux after 4--5 times the estimated lifetime of the initial load of rubidium assuming operation in the effusive regime.\cite{ramsey1956}

\section{\label{Pictures} Pictures}
\begin{figure}
\begin{center}
\includegraphics[width=15cm]{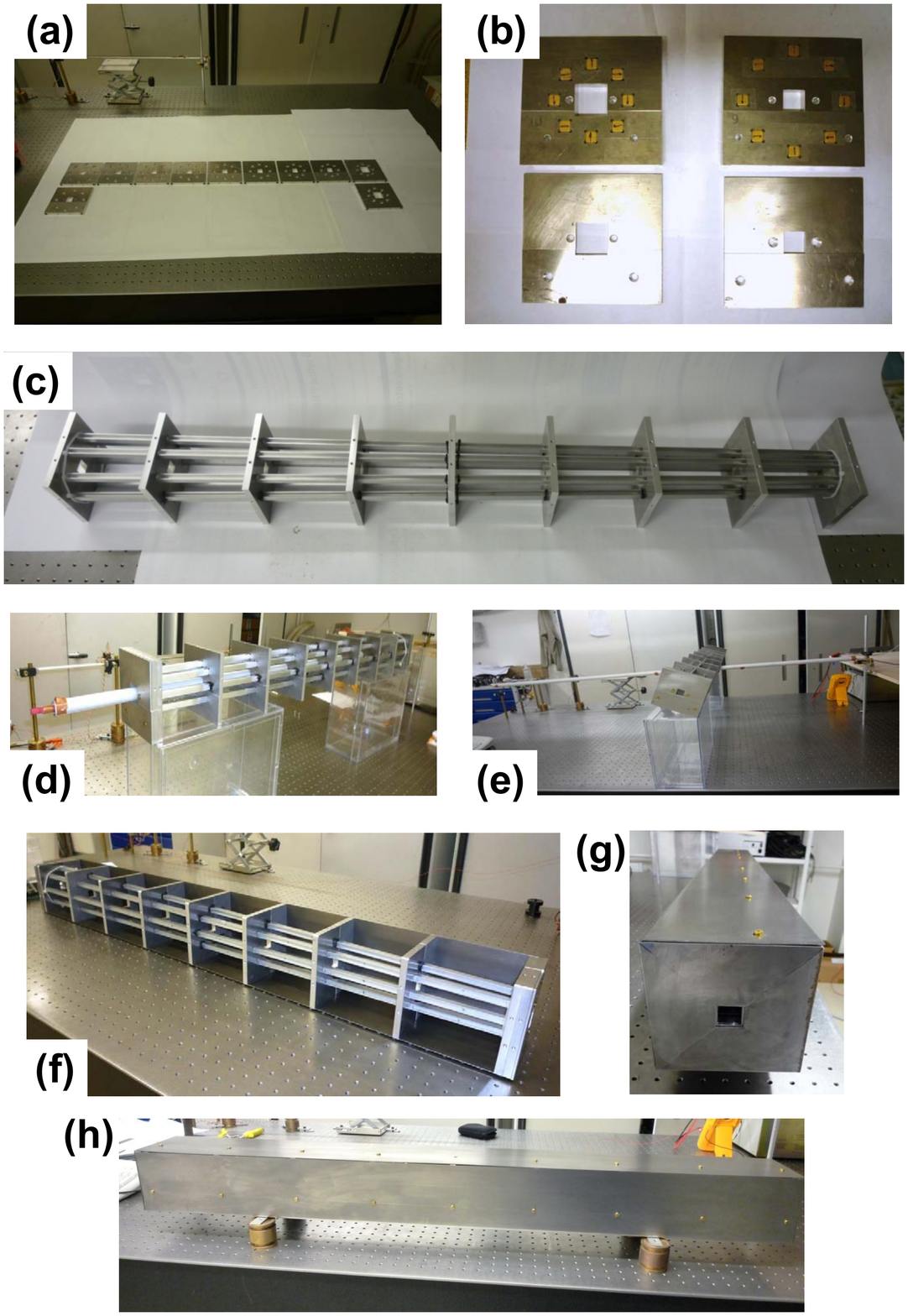}
\caption{(a) 9 mounts plus 2 endcaps. (b) top: endcaps with cube magnets inserted (golden squares)-arrows depicts magnetization direction; bottom: $5$~mm-thick spacers between endcaps and shield side (see picture (g) below). (c) implementation of the 9 mounts with the U-shaped profiles. Magnets will be inserted one by one in the grooves and clamped by a plastic wedge under each mount. (d) longitudinal B-field measurement. The pink 3D-probe is pulled into the gray plastic pipe at constant speed by wire attached to a stepper motor. Data is periodically recorded by a datalogger. (e) transverse B-field measurement. (f) back half part of the shield screwed on the mounts. (g) close view of the shield side. (h) assembly completed.}
\label{Fig:planche1}
\end{center}
\end{figure}

\begin{figure}
\begin{center}
\includegraphics[width=15cm]{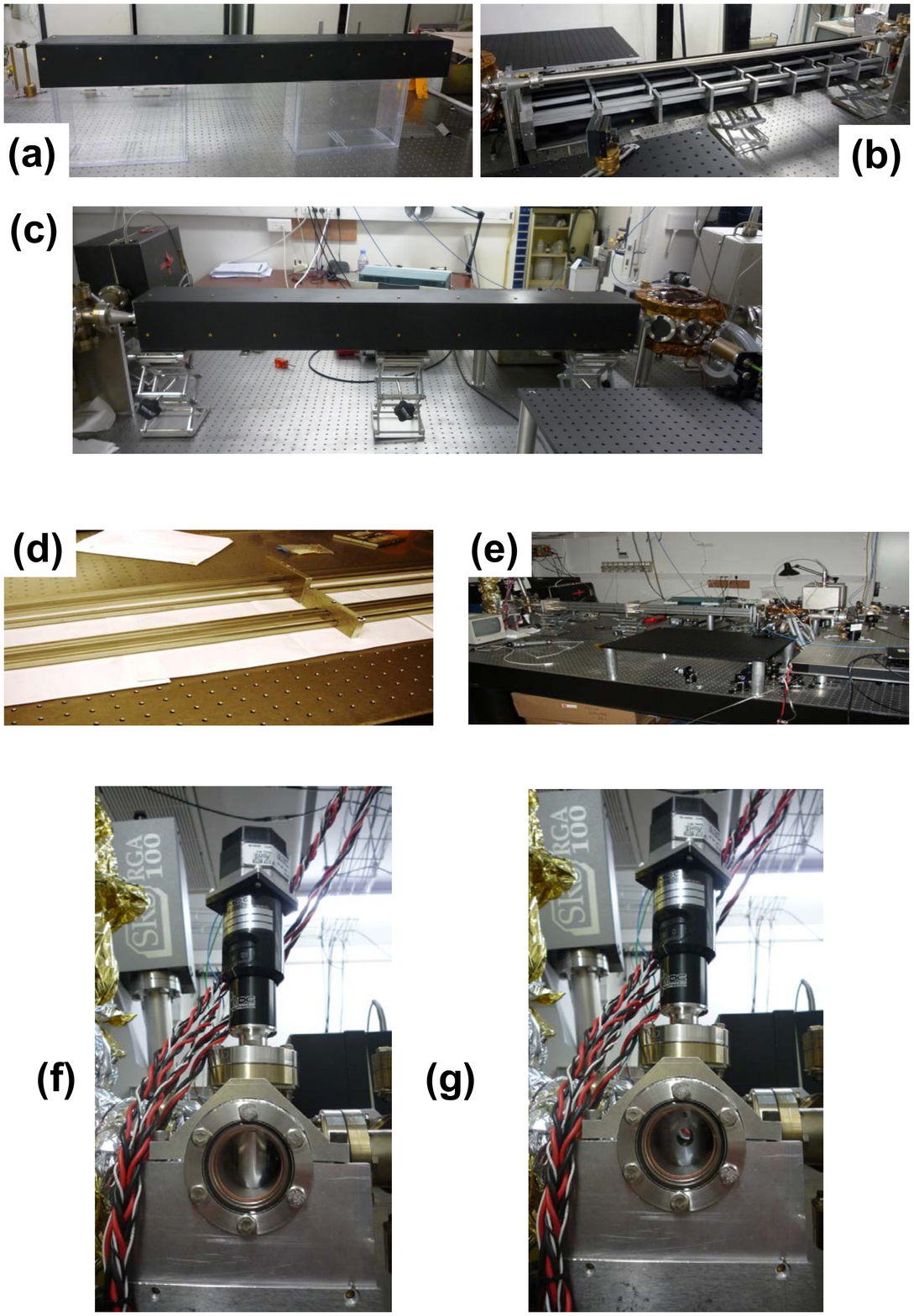}
\caption{(a) assembly completed with black paint. (b) the assembly is divided in two parts. The lower part is lifted and positioned below the CF16 UHV pipe. (c) The upper part of the assembly is screwed on the lower part and shield closed. In a few minutes, slow atoms will be delivered in the MOT chamber (right, with orange kapton tape for isolation of the MOT gradient coils). (d) $2^{nd}$ prototype with only three mounts: one at both ends, one in the middle. Early stage showing U-shaped profiles and the middle mount. (e) $2^{nd}$ prototype assembled around the CF16 pipe. Shield will be screwed later on. (f) beam shutter on. (g) beam shutter off.}
\label{Fig:planche2}
\end{center}
\end{figure}

We show next pages some pictures of different steps during the assembly of the Zeeman slower.
\begin{itemize}
  \item Fig.~\ref{Fig:planche1} and Fig.~\ref{Fig:planche2}~(a)-(c) are related to the setup described in the article.
  \item Fig.~\ref{Fig:planche2}~(d)-(e) are two pictures of a simplified setup with slightly higher field parameters $\Delta B=330\;\mathrm{G}$ and $B_{bias}=350\;\mathrm{G}$ corresponding to $d(0)=37.8\;\mathrm{mm}$ and $d(L)=27.2\;\mathrm{mm}$ and a slope of $-8.9\;\mathrm{mm/m}$. It was made recycling the magnets, the shield and two mounts from the original setup. A single new mount was machined. Conversion took less than one day work and it is the slower we have been using from then on.
  \item Fig.~\ref{Fig:planche2}~(f)-(g) are two pictures of our stepper-motor-actuated beam shutter.
\end{itemize}

\section{\label{Bibliography} Bibliography}

\end{document}